\documentclass[twocolumn,showpacs]{revtex4}

\usepackage{graphicx}
\usepackage{dcolumn}
\usepackage{amsmath}

\begin{document}

\title[Short Title]{Decay of spin-peierls state in CuGeO$_{3}$:Fe. The case of a strong disorder}

\author{S. V. Demishev}
 \email{demis@lt.gpi.ru}
 \affiliation{%
Low Temperatures Lab., General Physics Institute RAS,
 38 Vavilov str., 117942, Moscow, Russia}%
\affiliation{%
Venture Business Laboratory, Kobe University,
1-1 Rokkodai, Nada, Kobe 657-8501, Japan}%

\author{R.V.Bunting}
\author{A.A.Pronin}
\author{N.E.Sluchanko}
\author{N.A.Samarin}
\affiliation{%
Low Temperatures Lab., General Physics Institute RAS,
 38 Vavilov str., 117942, Moscow, Russia}%

\author{H.Ohta}
\author{S.Okubo}
\affiliation{%
Molecular Photoscience Research Center and Department of Physics, Kobe University,
1-1 Rokkodai, Nada, Kobe 657-8501, Japan}%

\author{Y.Oshima}
\affiliation{%
Graduate School of Science and technology, Kobe University,
1-1 Rokkodai, Nada, Kobe 657-8501, Japan}%

\author{L.I.Leonyuk\footnote{Deceased}}
\author{M.M.Markina}
\affiliation{%
Moscow State University, 119899 Moscow, Russia}%

\date{Received \today}

\begin{abstract}
Influence of doping by iron impurity on spin-Peierls state in CuGeO$_{3}$
is studied. ESR measurements for the frequency/temperature domain 60-450 GHz/
1.8-300 K and specific heat data obtained for the interval 6-20 K show that insertion
of 1\% of Fe completely destroy both spin-Peierls and antiferromagnetic orders. Damping of
long-range magnetic order is accompanied by onset at $T<$20 K of power asymptotics for
magnetic susceptibility $\chi$$\propto$$T^{-\alpha}$
and magnetic part of specific heat $c_{m}$$\propto$$T^{1-\alpha}$,
with the index $\alpha=$0.35-0.37.
This effect is characteristic to the limit of strong disorder
for doped CuGeO$_{3}$ and may reflect
formation of the Griffiths phase at low temperatures in CuGeO$_{3}$:Fe.
\end{abstract}


\maketitle

\subsection{Introduction}

Discovery of inorganic spin-Peierls compound CuGeO$_{3}$ opened
an opportunity to study
influence of doping and disorder on the spin-Peierls state.
Numerous experiments and
theoretical studies have been carried out in this field up to
now. However, from the theoretical point of view, most of
the available data correspond to the limit of weak disorder
when density of states have a pseudogap, i.e. a spin-Peierls
gap filled by disorder-induced states \cite{p1}. In this case the
expected temperature-concentration $T-x$ phase diagram consists
of uniform state, spin-Peierls state, antiferromagnetic state
and the region, where antiferromagnetic and spin-Peierls orders
coexist \cite{p1}. This structure of $T-x$ phase diagram was observed
experimentally for Zn, Si, Ni, Co, Mg and Mn impurities
\cite{p1,p2,p3,p4,p5,p6,p7,p8,p9,p10}.

In the limit of a strong disorder the ground state of CuGeO$_{3}$
is gapless and the density of states diverges at $\epsilon=$0:
$\rho$ ($\epsilon$) $\propto$$|${$\epsilon$}$|$$^{-\alpha}$ \cite{p1}.
As a consequence the temperature dependences of
magnetic susceptibility
$\chi$ and magnetic contribution
$c_{m}$ to specific heat $c_{p}$ acquire the forms \cite{p11}
\begin{equation}
 \label{eq:chi}
 \chi \propto T^{-\alpha}
\end{equation}
\begin{equation}
 \label{eq:cm}
 c_{m} \propto T^{1-\alpha}
\end{equation}
where $\alpha<1$. A non-Curie type behavior of $\chi(T)$ have been first
reported for CuGeO$_{3}$ doped with Zn \cite{p4}. However, experiments
in Ref.4 were carried out for extremely low Zn concentrations
corresponding to the \emph{weak} disorder limit, and the observed
deviations from the Curie law can not be related to the case of
strong disorder.

Recently Demishev et al. \cite{p12} suggested that CuGeO$_{3}$ doped
with Fe provides a "true" experimental realisation of a strongly
disordered regime \cite{p12}. It was found that substitution of Cu by
1\% of Fe in CuGeO$_{3}$ matrix induces strong disorder in magnetic
subsystem and leads to the complete damping of the spin-Peierls
transition. As long as the low temperature asymptotic of
the integrated intensity of the electron spin resonance (ESR)
line $I(T)$ corresponded to Eq. (1), it was concluded that doping with
iron gives rise to onset of a "quantum critical point" \cite{p12}
or, in the other words, a strongly disordered limit \cite{p1,p11}.

Nevertheless, the data presented in Ref.12 can not be considered as a complete evidence of
the aforementioned conclusion. Firstly, the ESR frequency 60 GHz
used in \cite{p12}
was high enough to cause possible violation
of the widely applied approximation $I(T)$$\propto$$\chi(T)$\cite{p13}. Indeed, in general case
the integrated intensity for arbitrary electromagnetic wave
frequency $\omega$ and resonant magnetic field $B_{res}$ is given by \cite{p14}
\begin{equation}
 \label{eq:IT}
 I(T) \propto \omega\frac{M(T,B_{res})}{B_{res}}
\end{equation}
and the observed in \cite{p12} deviations from Curie law may reflect
non-linearity of magnetic moment $M(T,B_{res})$ in strong magnetic field (note
that for $\omega/2\pi=$60 GHz the resonant field in CuGeO$_{3}$:Fe is about
$B_{res}\approx$ 2 T \cite{p12}).

Secondly, the iron concentration of 1\% is relatively low and
according to most findings should correspond to the weak
disorder region \cite{p1,p2,p3,p4,p5,p6,p7,p8,p9,p10}.
At the same time the absence of
antiferromagnetic resonance (AFMR) modes in \cite{p12} was confirmed
only up to $\omega/2\pi=$120 GHz. Therefore the possibility of coexistence
of ESR and AFMR, i.e. of spin-Peierls and antiferromagnetic
orders, in high frequency range $\omega/2\pi>$120 GHz for CuGeO$_{3}$:Fe
is not excluded. The described coexistence
is a fingerprint of considerable but still weak disorder \cite{p1},
so experiments at higher frequencies are highly desirable.

Thirdly, in the limit of a strong disorder Eqs. (1) and (2)
should be valid simultaneously. This have not been checked
in \cite{p12}, moreover as far as we know the information about
specific heat of the doped CuGeO$_{3}$ in this region is missing.

The aim of the present work is to solve experimentally
these three problems formulated above and verify interpretation
proposed in \cite{p12}. For that reason we studied single crystals
of Cu$_{0.99}$Fe$_{0.01}$GeO$_{3}$ obtained by self-flux technique \cite{p15}
and identical to the crystals studied in \cite{p12}. The quality
of crystals have been controlled by X-ray and Raman
scattering data; the actual contents of Fe in crystals
was determined by chemical analysis. The structure of
the samples studied coincided with the structure of pure
CuGeO$_{3}$ and the effect of doping on the Raman spectra
confirmed that iron impurity substitute cooper \cite{p12}.

Experimental facilities of three different kinds were used to study
magnetic and thermodynamic properties of CuGeO$_{3}$:Fe.
Magnetoabsorption lines for the frequencies up to 450
GHz were studied with the help of the magneto-optical facility at
Kobe University. In this experiment we measured transmission
through the sample as a function of magnetic field $B$ up
to 16 T at fixed frequency for liquid helium temperatures 4.2 K and 1.8 K.
Simultaneously a reference transmittance of
the thin layer of DPPH powder has been recorded and both magnetoabsorption
spectra of CuGeO$_{3}$:Fe and DPPH were analyzed quantitatively.

Temperature dependence of the ESR spectrum in the range
1.8-140 K was measured using 60 GHz cavity spectrometer
in General Physics Institute \cite{p12}. For each temperature
studied the accuracy of the temperature stabilization was better
than 0.01 K. All magnetoabsorption experiments reported below
were carried out in \textbf{B}$\parallel$\textbf{a} geometry.

Specific heat of CuGeO$_{3}$:Fe for the temperature interval
6-20 K was studied in Moscow State University with the help
of the low temperature small sample relaxation calorimeter.

\subsection{ESR studies of CuGeO$_{3}$:Fe}

Typical magnetoabsorption spectra in transmission
experiment are shown in Fig. 1. Up to $\omega/2\pi=$450 GHz neither
additional "impurity" lines, nor AFMR modes have been detected
and the spectrum for CuGeO$_{3}$:Fe consists of a single
lorenzian ESR line. The corresponding $g-$factor value is
close to that for Cu$^{2+}$ ions in the case
\textbf{B}$\parallel$\textbf{a}  (see below).
The resonant field for this line scales linearly with
frequency, i.e. $g$-factor is frequency independent
(see inset in Fig. 1).

\begin{figure}[tb]
\begin{center}
\includegraphics[width=\columnwidth]{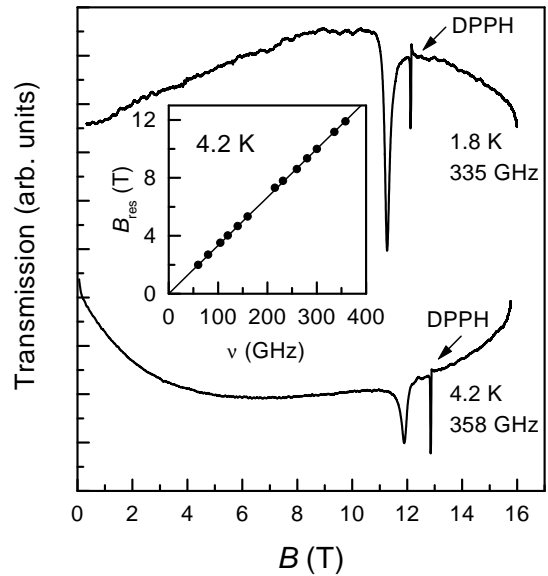}
\caption{Magnetoabsorption spectra for CuGeO$_{3}$:Fe in transmission experiment.}
\end{center}
\end{figure}

Transmission data obtained at various frequencies were used to
calculate ESR integrated intensities for CuGeO$_{3}$:Fe and DPPH.
It follows from Eq. (3) that for each frequency
\begin{equation}
 \label{eq:ii}
 \frac{I_{1}}{I_{0}}=\frac{M_{1}(T,B^{1}_{res})}{M_{0}(T,B^{0}_{res})}\cdot
 \frac{B^{0}_{res}}{B^{1}_{res}}
\end{equation}
where indexes 0 and 1 denote characteristics of ESR lines for DPPH and
CuGeO$_{3}$:Fe respectively. Assuming that magnetic moment of DPPH is given
by Brillouin function $M_{0}(T,B) \propto B_{J}(\mu_{B}B/k_{B}T)$
it is possible to calculate field or frequency
dependence of magnetic moment $M_{1}(T,B_{res})=f[B_{res}(\omega)]$
for CuGeO$_{3}$:Fe with the help of Eq. (4).
The result is presented in Fig. 2; it is visible that linear region
$M_{1} \propto B_{res}$ lasts up to $B_{res}\approx$6 T ($\omega/2\pi=$200 GHz).
At higher resonant fields/frequencies magnetic moment of CuGeO$_{3}$:Fe
tends to saturate and above $B_{res}\approx$11 T ($\omega/2\pi=$350 GHz)
$M_{1}(B_{res})$
starts to decrease with field (Fig. 2).

\begin{figure}[tb]
\begin{center}
\includegraphics[width=\columnwidth]{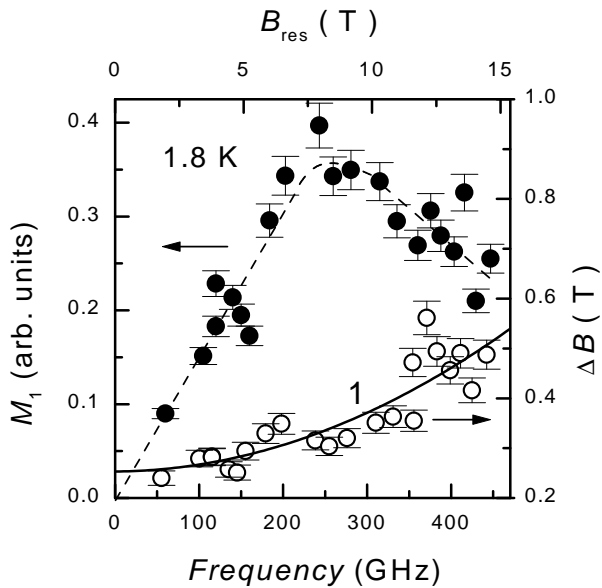}
\caption{Frequency and resonant field dependence of magnetic moment $M_{1}$ and ESR line
width $\Delta$$B$ for CuGeO$_{3}$:Fe at $T=$1.8 K. Curve 1 correspond to best fit of
$\Delta$$B(\omega)$ using Eq. (5).}
\end{center}
\end{figure}

Along with the integrated intensity the width $\Delta$$B$
of the ESR line in CuGeO$_{3}$:Fe was calculated. Contrary to the frequency independent
$g$-factor this parameter demonstrates a considerable frequency
dependence. It follows from Fig. 2 that $\Delta$$B$ increases two times
when frequency is varied from $\omega/2\pi=$60 GHz to $\omega/2\pi=$450 GHz.
Experimental data $\Delta$$B(\omega)$ at $T=$1.8 K can be modeled by expression
\begin{equation}
\label{eq:db}
\Delta B = A \cdot \omega^{2}+C
\end{equation}
which is characteristic to Raman relaxation mechanism for $S$=1/2 ion \cite{p16}.
The line 1 in Fig.2 correspond to best fit parameters in Eq. (5)
$A=$(3.2$\pm$0.4)$\cdot$10$^{-7}$T/GHz$^{2}$ and $C=$(0.063$\pm$0.004) T. It is
interesting, that attempts to model $\Delta$ $B(\omega)$ by expression
for the direct process $\Delta$$B$$\propto$$\omega^{5}$/tanh($\hbar\omega$/2k$_{B}$T)
\cite{p16} have failed as long as the theoretical frequency dependence was too
strong to fit the experimental data in Fig. 2. Therefore it is possible to
conclude that at low temperatures the dispersion of the relaxation time
in CuGeO$_{3}$:Fe is controlled mainly by the Raman process.

The above results indicate that at $\omega/2\pi=$60 GHz the
CuGeO$_{3}$:Fe remains in the region of linear magnetic response
and relation $I(T)$$\propto$$\chi$$(T)\equiv$$M(T,B_{res})/B_{res}$
is valid. In order to check assumptions of Ref. 12 we performed
also temperature measurements of the ESR in the 60 GHz cavity spectrometer
on the same crystal as was investigated in the quasi-optical transmission experiment.
For the precise determination of the $g$-factor a DPPH crystal was
placed in the cavity together with the CuGeO$_{3}$:Fe sample.

At all temperatures studied a single absorption line
of lorenzian shape was observed (Fig.3), that is in agreement with the results
of the quasi-optical experiment. Data in Fig.3 were used to calculate
temperature dependences of the $g$-factor $g(T)$, line width $\Delta$$B(T)$
and integrated intensity $I(T)$ (see Fig.4).

\begin{figure}[b]
\begin{center}
\includegraphics[width=\columnwidth]{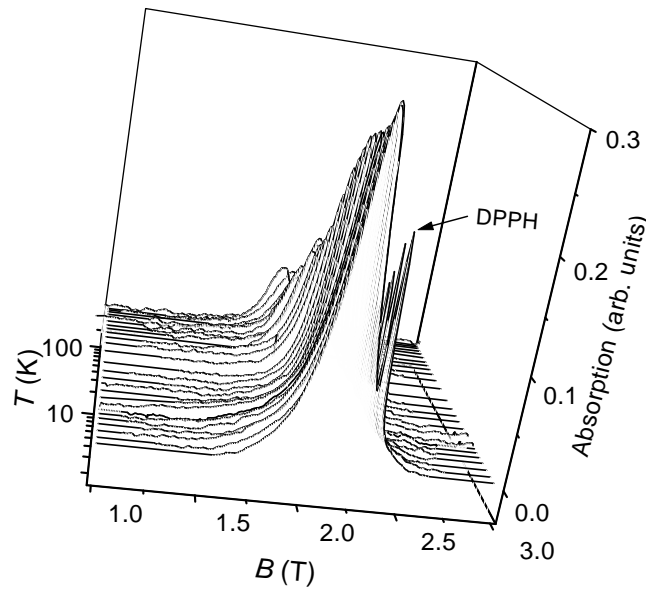}
\caption{Evolution of the ESR absorption line with temperature measured in cavity
spectrometer ($\omega/2\pi=$60 GHz, mode TE$_{011}$, quality factor $Q=$10$^{4}$).}
\end{center}
\end{figure}

For $T>$20 K the value of the $g$-factor is $g\approx$2.15 and characteristic
to Cu ions in CuGeO$_{3}$ structure for geometry $\mathbf{B}$$\parallel$$\mathbf{a}$
\cite{p17}. Below $T=$20 K $g$-factor starts to increase with lowering temperature and
reach the value $g=$2.19 at $T=$1.8 K (Fig. 4). It is worth to note, that
in the case of Fe-doped crystal no giant changes of the $g$-factor like in
Ni-doped CuGeO$_{3}$ \cite{p8} are observed.

\begin{figure}[tb]
\begin{center}
\includegraphics[width=\columnwidth]{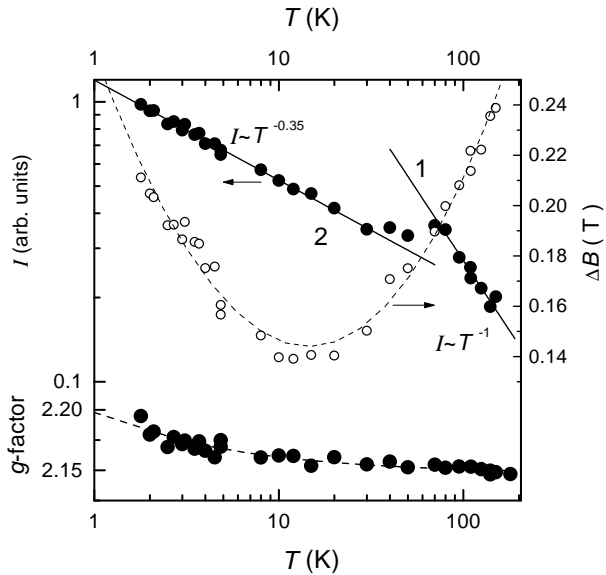}
\caption{Temperature dependences of the integrated intensity $I$, line width
$\Delta$$B$ and $g$-factor obtained in cavity experiment.}
\end{center}
\end{figure}

The temperature dependence of the line width is non-monotonic:
when temperature is lowered the $\Delta$$B(T)$ first decreases, passes through
a minimum at $T\approx$10-20 K, and finally starts to increase again.
It is interesting that in pure CuGeO$_{3}$ the width of the ESR line
decreases gradually with lowering temperature and the magnitude of $\Delta$$B(T)$
at $T=$100 K is about 6 times smaller than in Fe-doped crystal \cite{p17}.
The decrease of temperature makes a difference in $\Delta$$B(T)$ more dramatic:
at $T=$1.8 K the line width for the Fe-doped CuGeO$_{3}$ is 200 times bigger
than in pure crystal (see Fig. 4 and data from Ref.17).

The significant difference between pure and Fe-doped CuGeO$_{3}$ is visible
in the temperature dependence of the integrated intensity. It follows from
Fig. 4 that in the Fe-doped crystal spin-Peierls transition is completely
damped. For $T>$70 K integrated intensity obeys Curie law $I(T)\propto$$T^{-1}$
(Fig. 4, curve 1). In the temperature range 25-70 K $I(T)$ saturates, and at
lower temperatures a power law asymptotic behavior $I(T)\propto$$T^{-\alpha}$ with
the index $\alpha=$0.35$\pm$0.03 is observed (Fig. 4, curve 2). This result
agrees well with the data obtained in Ref.12.

\subsection{Specific heat in CuGeO$_{3}$:Fe}

The experimental limitation on the sample mass in
specific heat measurements has not allowed to study
individual crystals as in ESR experiments and a set of
CuGeO$_{3}$:Fe single crystals cleaved from the same ingot and having
total mass about 40 mg have been used. The temperature dependence
of specific heat $c_{p}(T)$ for Fe-doped and pure CuGeO$_{3}$ is
presented in Fig. 5 (curves 1 and 4 respectively). The sharp peak
at spin-Peierls transition in CuGeO$_{3}$:Fe sample have vanished,
that is in agreement with the ESR evidence of complete damping
of the spin-Peierls transition by iron impurity (Fig. 4).

From the fact that in Fe-doped crystal specific heat is considerably
(about 50\% for $T>$14 K) bigger than in pure CuGeO$_{3}$ (Fig. 5)
it is reasonable to suppose that the studied sample has an excessive
magnetic contribution $c_{m}(T)$ in $c_{p}(T)$:
\begin{equation}
 \label{eq:cp}
 c_{p}(T)=c_{D}(T)+c_{m}(T)= \beta T^{3} + \gamma T^{\delta}
\end{equation}
Here the first term represents lattice (Debye) part and the choice
for the analytical representation of the magnetic term
$c_{m}$(T)=$\gamma$$T^{\delta}$ is made
in accordance with Eq. (2).

\begin{figure}[b]
\begin{center}
\includegraphics[width=\columnwidth]{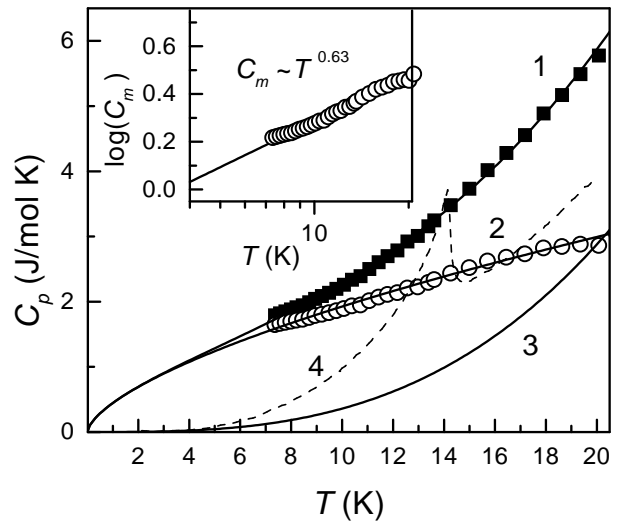}
\caption{Temperature dependence of specific heat in CuGeO$_{3}$:Fe.
Curve 1: points- experiment, line- best fit using Eq. (6). Curve 2: points-
experimental data for magnetic specific heat obtained from $c_{p}(T)$ by substraction
of the lattice contribution (curve 3); line- best fit of magnetic part.
Curve 4- specific heat for pure CuGeO$_{3}$ (from \cite{p18}).}
\end{center}
\end{figure}

The Eq. (6) was applied to model experimental data for CuGeO$_{3}$:Fe (Fig. 5).
We find the value $\beta=$(0.36$\pm$0.02) mJ$\cdot$mol/K, corresponding to
Debye temperature $\Theta_{D}\approx$300 K. This result agrees well
with the previous findings for pure CuGeO$_{3}$ \cite{p18}, where
$\Theta_{D}=$310 K have been obtained. Consequently the main
changes in specific heat caused by doping with iron occur in magnetic part
$c_{m}(T)$ (see curve 2 in Fig. 5; the magnetic contribution is obtained by
subtracting of the lattice part from $c_{p}$$(T)$). The best fit of $c_{m}$$(T)$
gives parameters values $\delta=$0.63$\pm$0.04 and $\gamma=$(0.44$\pm$0.04)
mJ$\cdot$mol/K$^{1+\delta}$. The power law for magnetic part of specific
heat is illustrated by inset in Fig. 5.

\subsection{Summary and discussion}

Summarizing experimental results of the present work, we wish
to mark that observed low temperature behavior of the ESR line reflects
intrinsic properties of Cu$^{2+}$ chains modified by Fe impurity
rather than impurity paramagnetism caused by Fe ions.
This conclusion can be deduced from the $g$-factor values characteristic to Cu$^{2+}$
(Fig. 4) and the observation of the line width frequency dependence given
by Eq. (5). Indeed, for the Fe$^{2+}$ ion substituting Cu$^{2+}$ ion in $S=$1/2
chain a spin state with $S=$2 may be expected \cite{p19}. For the integer
impurity spin the term proportional to $\omega^{2}$ in expression
for the line width should vanish and $\Delta B(T)$ will be frequency
independent \cite{p16}. The expected "impurity behavior" for Fe$^{2+}$
contradicts to experimental data (Fig. 2) and the whole
experimental picture of ESR in CuGeO$_{3}$:Fe is consistent
with the assumption that magnetic properties of this system
are controlled by disordered Cu$^{2+}$ chains. The presence
of the disorder in magnetic subsystem follows from
the strong broadening of the ESR line with respect to the
pure crystal (Fig. 4 and Ref.17) and agrees with the results
of the structural studies \cite{p12}.

As long as the integrated intensity measured at $\omega/2\pi=$60 GHz for CuGeO$_{3}$:Fe
is proportional to magnetic susceptibility, the latter quantity
diverges at $T<$20 K: $\chi(T)\propto$$T^{-\alpha}$, where $\alpha\approx$0.35.
According to Eq. (2) the value of the index $\delta$ in Eq(6)
should be $\delta=1-\alpha\approx$0.65, whereas experiment
gives $\delta=$0.63 (Fig. 5). Both values of the indexes are coincide within the
experimental error and therefore Eq. (1) and Eq. (2) are hold simultaneously. This
result confirms that inserting of 1\% of iron in CuGeO$_{3}$ matrix really
induces a strongly disordered limit of doping as it was proposed in \cite{p12}.

From the theoretical point of view the non-Curie asymptotic behavior of magnetic
susceptibility and related power law for magnetic part of specific heat
reflect the onset of the Griffiths phase (GP) which thermodynamic
properties are controlled by relatively rare spin clusters correlated
more strongly than average [20-22]. The GP appears in various spin
systems below some critical temperature $T_{G}$ if the magnitude of random
potential is strong enough to destroy transition to magnetically ordered
phase [20-22] (in the case of CuGeO$_{3}$:Fe we have shown that both spin-Peierls
and antiferromagnetic transitions are completely damped by doping).
According to the data in Fig. 4 the value of $T_{G}$ in CuGeO$_{3}$:Fe
can be estimated as $T_{G}$=20-40 K. The possible interaction
effects inside spin clusters of GP may be also responsible
for a weak temperature dependence of the $g$-factor observed at $T<T_{G}$
(Fig. 4) and unusual field dependence of magnetic moment (Fig. 2).
However this problem requires further theoretical investigation.

It is interesting, that 1\% of iron is sufficient to shift doped CuGeO$_{3}$
into the region of the strong disorder, whereas another impurities studied
\cite{p1,p2,p3,p4,p5,p6,p7,p8,p9,p10} have corresponded to
a weak disorder limit \cite{p1}.
This question deserves a detail discussion of the universal properties of
$T-x$ phase diagram of CuGeO$_{3}$ and is reserved for future publications.

\begin{acknowledgments}
Authors are grateful to A.N.Vasil'ev for valuable discussions and V.V.Glushkov
for assistance. This work was
supported by programmes "Physics of Microwaves" and "Fundamental Spectroscopy"
of Russian Ministry of Industry, Science and Technology. SVD acknowledge
financial support from Venture Business Laboratory in Kobe University.
\\
\end{acknowledgments}

\end{document}